\documentclass[11pt ]{revtex4}
\usepackage{hyperref}
\usepackage{amsmath,amsfonts}
\usepackage[dvips]{graphics}
\usepackage[dvips]{graphicx}
\usepackage{subfigure}
\usepackage{epsfig}
\usepackage{amsmath}
\usepackage{amsfonts}
\RequirePackage{color}

\begin{document}

\title{Thermodynamics of Higher Order Entropy Corrected Schwarzschild-Beltrami-de Sitter Black Hole}
\author{B. Pourhassan}
 \email{b.pourhassan@du.ac.ir}
  \affiliation{School of Physics, Damghan University, Damghan, Iran}
\author{S. Upadhyay}
   \email{sudhakerupadhyay@gmail.com, sudhaker@associates.iucaa.in}
  \affiliation{Department of Physics, K. L. S. College,  Nawada-805110,  India}
\affiliation{Visiting Associate, Inter-University Centre for Astronomy and Astrophysics (IUCAA) Pune, Maharashtra-411007}
 
\author{H. Farahani}
 \email{h.farahani@umz.ac.ir}
  \affiliation{School of Physics, Damghan University, Damghan, Iran}

\begin{abstract}
 In this paper, we consider higher order correction of the entropy and study the thermodynamical properties of recently proposed Schwarzschild-Beltrami-de Sitter black hole, which is indeed an exact solution of Einstein equation with a positive cosmological constant. By using the corrected entropy and Hawking temperature we extract some thermodynamical quantities like Gibbs and Helmholtz free energies and heat capacity. We also investigate the first and second laws of thermodynamics. We find that   presence of higher order corrections, which come from thermal fluctuations, may remove some instabilities of the black hole. Also unstable to stable phase transition is possible in presence of the first and second order corrections. 
\end{abstract}

\maketitle

\section{Introduction}
As we know, the maximum entropy of black holes is proportional to the event horizon area \cite{2, 4}. It yields to the holographic principle development \cite{5, 5a}. However, mentioned maximum entropy of the black holes will be corrected due to quantum fluctuations of statistical mechanics, which yields to modify the holographic principle \cite{6, 6a}. These are indeed thermal fluctuations around equilibrium and become important as the black hole size reduced due to
Hawking radiation. In that case, quantum fluctuations will correct the well-known relation between the entropy and the black hole area. These corrections have been evaluated using several different approaches.

Das et al. \cite{7} found that leading-order correction of the entropy of any thermodynamic system due to small statistical fluctuations around equilibrium is logarithmic. Their result examined for BTZ black holes, anti-de Sitter Schwarzschild and Reissner-Nordstrom black holes in arbitrary dimensions. Such quantum correction applied to several black holes like recently proposed three-dimensional hairy black hole \cite{9, 10, 11, 12, 13} and found that it is logarithmic at the first order approximation \cite{8}. In that case the effects of the logarithmic correction on a charged  black hole in anti de Sitter space has been studied \cite{14}. Recently, quantum corrections to the thermodynamics of quasitopological black holes
 \cite{sudu}, a charged rotating AdS black holes \cite{sud18}, a charged black holes in gravity's rainbow \cite{sud19}, a static black hole in $f(R)$ gravity \cite{sud20} and Ho\v rava-Lifshitz black holes \cite{sudu1} due to thermal fluctuation are discussed. Also, modified thermodynamics and statistics of G\"{o}del black hole due to the logarithmic correction investigated \cite{15}. Moreover, the effects of the logarithmic correction on the thermodynamics of a modified Hayward black hole has been studied and found that this does not affect the stability of the black hole \cite{18}. The logarithmic correction effect has been studied for several black objects like black Saturn \cite{16} or charged dilatonic black Saturn \cite{17} and found that thermal fluctuations have important consequences. Logarithmic correction together AdS/CFT correspondence yield to interesting results, for example P-V criticality of charged AdS space-time \cite{momen} has been studied, or dyonic charged AdS black hole has been considered \cite{19} and found that this is holographic dual of a Van der Waals fluid \cite{20} under effect of the logarithmic correction.  The logarithmic corrected Van der Waals black holes in higher dimensional AdS space is also investigated recently \cite{sud0}.
 The P-V criticality of first-order entropy corrected AdS black holes in massive gravity is also studied \cite{sud1}.
In the recent work, quantum gravitational effect has been studied by using Dumb holes \cite{Annals}.

Quantum corrections to the Bekenstein-Hawking entropy of charged black hole using generalized uncertainty principle has been studied \cite{GUP}, which is applied to Schwarzschild-Tangherlini black hole \cite{ST1}. 
As mentioned above, the logarithmic correction is the first order approximation correction to the black hole entropy, while it is possible to calculate all orders of corrections to the entropy of black holes due to the statistical fluctuations around the equilibrium \cite{sur}. In this paper we would like to use this result to study modified thermodynamics of Schwarzschild-Beltrami-de Sitter black hole \cite{SBdS} which is indeed an unstable dynamical black hole and obtained by introducing inertial Beltrami coordinates to traditional non-inertial Schwarzschild-de Sitter metric.

It is argued that results of \cite{sur} are applicable to all classical black holes at thermodynamics equilibrium. However, under the special conditions one can use the first and second order of corrections in some unstable black holes. In that case the approximation is valid as one consider only small fluctuations near equilibrium. But we should comment that as the black hole becomes really small possible near Planck scale such approximation should break down and we have to use non-equilibrium methods to analyze the system. It may be dominant of quantum black holes. But as long as we consider small fluctuations we can analyze them perturbatively, and therefore we consider only the first and second order corrections and neglect the higher order terms in the perturbative expansion.

In the Ref. \cite{ha} authors investigated the thermodynamical properties of Schwarzschild-Beltrami-de Sitter black
hole and obtained some thermodynamics quantities like entropy, Hawking temperature, Gibbs free energy and heat capacity. They also investigate the
Smarr relations and the first law of thermodynamics. It is found that the Schwarzschild-Beltrami-de Sitter
black hole is unstable without ant phase transition and critical points as well as Schwarzschild-de Sitter black hole. Now, in this paper we consider higher order corrected entropy (to the second order) and show that Schwarzschild-Beltrami-de Sitter
black hole may be stable at allowed regions of event horizon radius.

This paper is organized as follows. In the next section we review some important aspect of Schwarzschild-Beltrami-de Sitter black hole and in section 3 we introduce corrections of the entropy. In section 4 we begin thermodynamical analysis under effect of thermal fluctuations and in section 5 calculate Gibbs and Helmholtz free energies. In section 6 we obtain heat capacity and discuss about stability of black hole. Finally in section 7 we give conclusion and summary of results.

\section{Schwarzschild-Beltrami-de Sitter Black Hole}
Schwarzschild-Beltrami-de Sitter black hole given by the following metric \cite{SBdS},
\begin{eqnarray}\label{metric}
ds^{2} &=& \left(\frac{1-\frac{r^{2}}{\sigma l^{2}}-\frac{2M\sqrt{\sigma}}{r}}{(1-\frac{t^{2}}{l^{2}})^{2}}-\frac{r^{2}t^{2}}{l^{4}(1-\frac{r^{2}}{\sigma l^{2}}-\frac{2M\sqrt{\sigma}}{r})\sigma^{3}}\right)dt^{2}-\frac{(1-\frac{t^{2}}{l^{2}})^{2}}{(1-\frac{r^{2}}{\sigma l^{2}}-\frac{2M\sqrt{\sigma}}{r})\sigma^{3}}dr^{2}\nonumber\\
&-&2\frac{rt(1-\frac{t^{2}}{l^{2}})}{l^{2}(1-\frac{r^{2}}{\sigma l^{2}}-\frac{2M\sqrt{\sigma}}{r})\sigma^{3}}dtdr-\frac{r^{2}}{\sigma}(d\theta^{2}+\sin^{2}\theta d\phi^{2}),
\end{eqnarray}
where $M$ is the mass and cosmological radius $l$ related to the cosmological constant via $l^{2}=3/\Lambda$, also $\sigma$ is defined by
\begin{equation}\label{sigma}
\sigma=1+\frac{r_i^2-t^2}{l^2},
\end{equation}
where $r_{i}$ is horizon radius which is obtained using the following equation,
\begin{equation}\label{horizon-equation}
1-\frac{r^{2}}{\sigma l^{2}}-\frac{2M\sqrt{\sigma}}{r}=0,
\end{equation}
which has three following solutions \cite{ha},
\begin{eqnarray}\label{horizon-solution}
r_{1}^{2}&=& \frac{4(l^{2}-t^{2})\sin^{2}\left(\frac{1}{3}\sin^{-1}(\frac{3\sqrt{3}M}{l})\right)}{1+2\cos\left(\frac{2}{3}\sin^{-1}(\frac{3\sqrt{3}M}{l})\right)},\nonumber\\
r_{2}^{2}&=&\frac{4(l^{2}-t^{2})\cos^{2}\left(\frac{1}{3}\sin^{-1}(\frac{3\sqrt{3}M}{l})+\frac{\pi}{6}\right)}
{3-4\cos^{2}\left(\frac{1}{3}\sin^{-1}(\frac{3\sqrt{3}M}{l})+\frac{\pi}{6}\right)} \nonumber\\
r_{3}^{2}&=&\frac{4(l^{2}-t^{2})\cos^{2}\left(-\frac{1}{3}\sin^{-1}(\frac{3\sqrt{3}M}{l})+\frac{\pi}{6}\right)}
{3-4\cos^{2}\left(\frac{1}{3}\sin^{-1}(-\frac{3\sqrt{3}M}{l})+\frac{\pi}{6}\right)}
\end{eqnarray}
The first root $r_{1}$, which is positive real root of the equation (\ref{horizon-equation}), is event horizon radius which we will denote by $r_{h}$ and can be rewritten as follow,
\begin{equation}\label{horizon-solution-2}
r_{1}\equiv r_{h}=\sqrt{\frac{(l^{2}+t^{2})(X^{\frac{4}{3}}+9l^{4})}{X^{\frac{4}{3}}+9l^{2}X^{\frac{2}{3}}+9l^{4}}},
\end{equation}
where we used the following definition,
\begin{equation}\label{defX}
X=-27Ml^{2}+3l^{2}\sqrt{81M^{2}-3l^{2}},
\end{equation}
which tells that,
\begin{equation}\label{condition-horizon}
M\geq\frac{l}{3\sqrt{3}},
\end{equation}
in agreement with the Ref. \cite{ha}. We will see that time-dependence of horizon radius don't yields any problem to study thermodynamics because horizon area will be independent of time. On the other hand we should note that dynamics of the Schwarzschild-Beltrami-de Sitter black hole described by cosmological constant with infinitesimal value which has no important effect. Hence, we can approximately considered the Schwarzschild-Beltrami-de Sitter black hole as static space-time.
\section{Thermal Fluctuations}
In this section we calculate the effect of thermal fluctuations on
the entropy of Schwarzschild-Beltrami-de Sitter black hole. In the seminal paper by Hawking and Page \cite{hawp}, it is shown that black holes in asymptotic curved space-time follow canonical ensemble. Based on this, the system of Schwarzschild-Beltrami-de Sitter black holes is considered as a canonical ensemble consisting $N$ particles with energy spectrum $E_{n}$. One can write
the statistical partition function of the system as,
\begin{equation}\label{partition}
Z = \int_0^\infty  dE  \rho (E) e^{-\beta_{\kappa} E},
\end{equation}
where $\beta_{\kappa}$ is the inverse of the temperature in unit of the Boltzmann constant, and $\rho (E)$ is the canonical density of the system with energy average $E$.
The partition function (\ref{partition}) together Laplace inversion can be used to calculate the density of states,
\begin{eqnarray}\label{density-state}
\rho (E) = \frac{1}{2 \pi i} \int^{\beta_{0\kappa}+
i\infty}_{\beta_{0\kappa} - i\infty} d \beta_{\kappa}  e
^{S(\beta_{\kappa})} ,
\end{eqnarray}
where
\begin{equation}
S = \beta_{\kappa}  E   + \log Z.
\end{equation}
The  entropy  around the equilibrium temperature $\beta_{0\kappa}$
will be obtained by eliminating all the thermal fluctuations.
However, in presence of the  thermal fluctuations, corrected entropy can be written by Taylor expansion around the $\beta_{0\kappa}$ as,
\begin{equation}\label{fluc}
S = S_0 + \frac{1}{2}(\beta_{\kappa} - \beta_{0\kappa})^2 \left(\frac{\partial^2
S(\beta_{\kappa})}{\partial \beta_{\kappa}^2 }\right)_{\beta_{\kappa} = \beta_{0\kappa}}+ \frac{1}{6}(\beta_{\kappa} - \beta_{0\kappa})^3 \left(\frac{\partial^3
S(\beta_{\kappa})}{\partial \beta_{\kappa}^3 }\right)_{\beta_{\kappa} = \beta_{0\kappa}}+\cdots,
\end{equation}
where dots denote higher order corrections. We should note that the first derivative of the entropy with respect to $\beta_{\kappa}$ vanishes at the equilibrium temperature.
So, the density of states can be written as
\begin{eqnarray}
\rho (E) &=& \frac{e^{S_0}}{ 2 \pi i}  \int^{\beta_{0\kappa}+
i\infty}_{\beta_{0\kappa} - i\infty}  d \beta_{\kappa} \, \,  \exp \left(
\frac{(\beta_{\kappa}- \beta_{0\kappa})^2}{2} \left(\frac{\partial^2
S(\beta_{\kappa})}{\partial \beta_{\kappa}^2 }\right)_{\beta_{\kappa} = \beta_{0\kappa}}
\right.\nonumber\\
& +&\left.\frac{(\beta_{\kappa}- \beta_{0\kappa})^3}{6} \left(\frac{\partial^3
S(\beta_{\kappa})}{\partial \beta_{\kappa}^3 }\right)_{\beta_{\kappa} = \beta_{0\kappa}}+\cdots  \right).\nonumber\\
\end{eqnarray}
In that case, following the Ref. \cite{sur} one can obtain,
\begin{equation}
S = S_0 -\frac{1}{2} \log{S_{0}T_{h}^{2}}+\frac{f(m,n)}{S_{0}}+\cdots,
\end{equation}
where $f(m,n)$ can be considered as a constant. However, we can write a general expression for the corrected entropy as \cite{PF},
\begin{eqnarray}\label{corrected-entropy}
S=S_0-\frac{\alpha_1}{2}\log(S_0T_{H}^2)+\frac{\alpha_2}{S_0}+...,
\end{eqnarray}
where we introduced a parameter $\alpha_1$ by hand to track corrected terms, so in the limit $\alpha_1\rightarrow0$, the original results can
be recovered and $\alpha_1=1$ yields usual corrections \cite{sur,SPR}. Hence the first order correction is logarithmic, while the second order correction is proportional to the inverse of original entropy $S_{0}$. These corrections can be considered as quantum correction of the black hole. One can neglect these corrections for the large black holes. However, by decreasing size of black hole due to the Hawking radiation, the quantum fluctuations in the geometry of the black hole increased. Thus, the thermal fluctuations modify the thermodynamics of the black holes and will be more important when the black holes reduced in size. In the next section we begin to analyze the thermodynamics quantities of the black hole with the corrected entropy.

\section{Thermodynamical Analysis}
\begin{figure}[h!]
 \begin{center}$
 \begin{array}{cccc}
\includegraphics[width=70 mm]{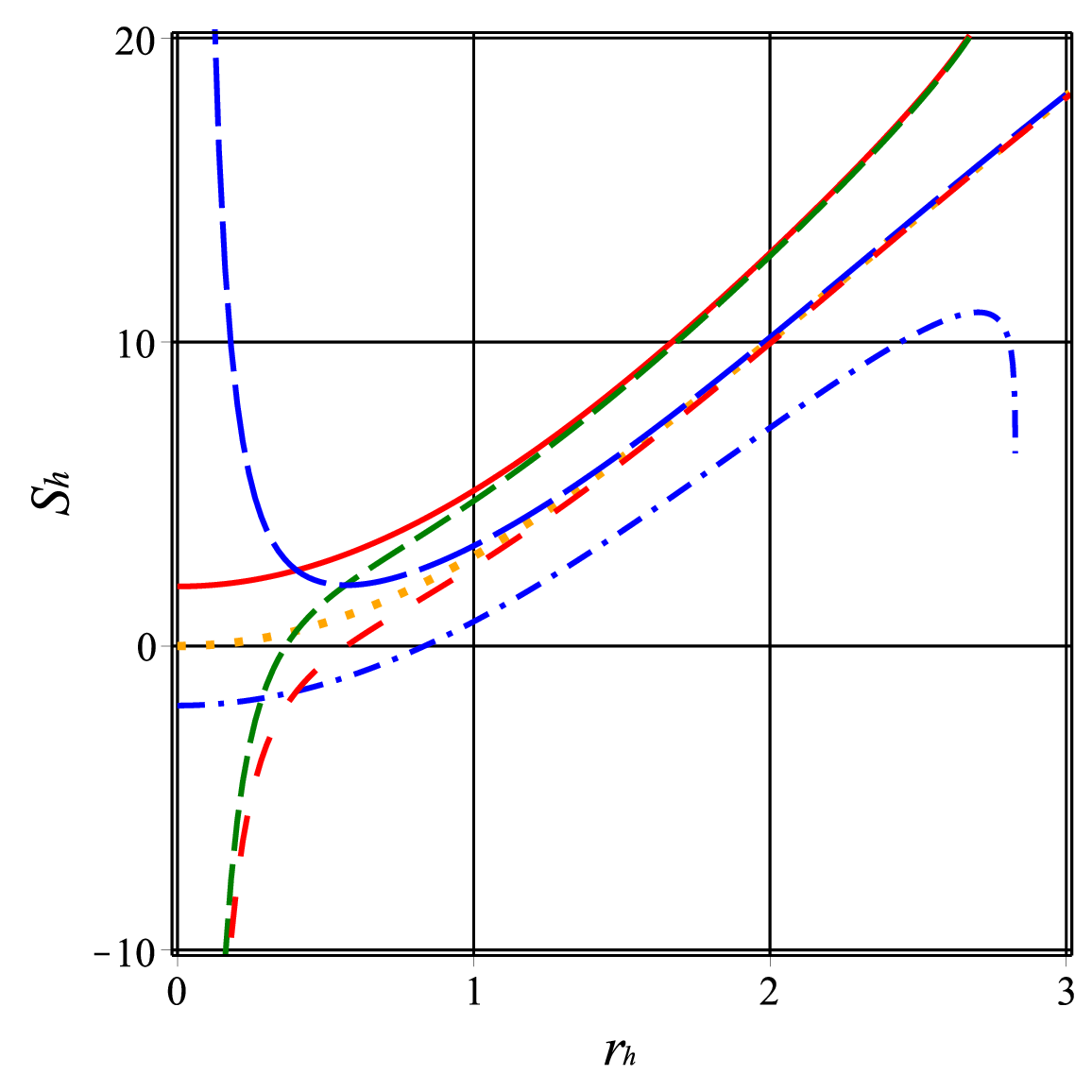}
 \end{array}$
 \end{center}
\caption{Entropy in terms of $r_{h}$ with $l=4$ and $t=0$. $\alpha_{1}=\alpha_{2}=0$ (dot orange), $\alpha_{1}=1$, $\alpha_{2}=0$ (solid red), $\alpha_{1}=0$, $\alpha_{2}=-1$ (space dash red), $\alpha_{1}=0$, $\alpha_{2}=1$ (long dash blue), $\alpha_{1}=-1$, $\alpha_{2}=0$ (dash dot blue).}
 \label{fig1}
\end{figure}

Thermodynamics of Schwarzschild-Beltrami-de Sitter black hole has been discussed in Ref. \cite{ha}. It has been argued that Bekenstein-Hawking area entropy $S_{0}=A/4$ is applicable because Schwarzschild-Beltrami-de Sitter black hole can be approximately considered as static space-time.  This consideration is based on discussion of Ref. \cite{hey}  which states that entropy-area relation for
black holes in stationary space-time is also applicable to dynamical black hole.  
Also, the time dependence of Schwarzschild-Beltrami-de Sitter metric is
attributed to the existence of cosmological constant as Schwarzschild-Beltrami-de Sitter space-time
converts to Schwarzschild  (static) space-time when cosmological radius tends to infinity, i.e.
cosmological constant tends to zero. According to the present standard model of cosmology,     the relevant  order of the cosmological constant is $10^{-52}$
in metric units, which of course can not influence 
the time independence of the metric of the original Schwarzschild space-time significantly.  Therefore, the time $t$ in the metric can be considered as a parameter rather than a dynamical variable. Hence, we can use general relation for the horizon area of spherically symmetric and static four-dimensional black hole,
\begin{eqnarray}\label{Area}
A_i=\int{d\theta d\phi \sqrt{g_{\theta\theta}g_{\phi\phi}}},
\end{eqnarray}
where $i=1,2,3$ corresponding to three horizons. Hence, one can obtain,
\begin{eqnarray}\label{ent}
S_0=\pi x_i^2,
\end{eqnarray}
with $x_i=r_i/\sqrt{\sigma}$.
So entropy $S_0$ in terms of $r_i$ reads,
\begin{eqnarray}
S_0=\pi \frac{r_i^2}{\sigma},
\end{eqnarray}
Then, the temperature is given by \cite{ha},
\begin{eqnarray}\label{tem}
T_i=\frac{1}{4\pi}\frac{l^2-3x_i^2}{l^2 x_i}=\frac{l^2-2r_i^2-t^2}{4\pi lr_i\sqrt{l^2+r_i^2-t^2}},
\end{eqnarray}
which   yields to the Schwarzschild black hole temperature at $l\rightarrow\infty$ limit. It is clear that the temperature is decreasing function of horizon radius, so when the size of black hole decreased, the temperature grow up and thermal fluctuations will be important as mentioned in the introduction, hence we should take into account such effects.\\
Also, the thermodynamical volume $V_i$ is computed as \label{ha}
\begin{eqnarray}\label{vol}
V_i =\frac{4}{3}\pi x_i^3.
\end{eqnarray}
One can also write the black hole mass using the Smarr formula as,
\begin{eqnarray}
M&=&\frac{x_i(l^2-3x_i^2)}{2l^2} +\frac{x_i^3}{l^2}-\alpha_1\frac{l^2-3x_i^2}{2\pi l^2 x_i}\log \frac{l^2-3x_i^2}{\sqrt{\pi} l^2} +\alpha_1 \left(\frac{\log 2}{\pi}\right)\frac{l^2-3x_i^2}{l^2x_i}\nonumber\\
&+&\frac{\alpha_2}{2\pi^2}\frac{l^2-3x_i^2}{l^2x_i^3}.
\end{eqnarray}
This   further simplifies to
\begin{eqnarray}\label{mass}
M&=& \frac{r_hl(l^2-2r_h^2-t^2)}{2(l^2+r_h^2-t^2)^{\frac{3}{2}}}+\frac{r_h^3 l}{(l^2+r_h^2-t^2)^{\frac{3}{2}}} -\alpha_1\frac{(l^2-2r_h^2-t^2)}{2\pi lr_h\sqrt{l^2+r_h^2-t^2}}
\log \frac{l^2-2r_h^2-t^2}{\sqrt{\pi}(l^2+r_h^2-t^2)}\nonumber\\
&+& \alpha_1\frac{\log 2}{\pi}\frac{(l^2-2r_h^2-t^2)}{lr_h\sqrt{l^2+r_h^2-t^2}}+
 \frac{\alpha_2}{2\pi^2}\frac{(l^2-2r_h^2-t^2)(l^2+r_h^2-t^2)^\frac{1}{2}}{r_h^3l^3}.
\end{eqnarray}
Above relations satisfy the first law of thermodynamics,
\begin{equation}\label{first}
dM=TdS+VdP,
\end{equation}
where the positive cosmological constant is related to the negative pressure
\begin{equation}\label{pr}
P=-\frac{\Lambda}{8\pi},
\end{equation}
with $\Lambda=3/l^2$. Hence, one can image $dP=0$ and the equation (\ref{first}) is valid for the original entropy given by (\ref{ent}) in the following region,
\begin{equation}\label{bound0}
0<r_{h}\leq\sqrt{\frac{l^{2}-t^{2}}{2}}.
\end{equation}
Also, by analyzing of heat capacity, in the region given by (\ref{bound0}), it has been found that the Schwarzschild-Beltrami-de Sitter black hole as well as Schwarzschild-de Sitter black hole are in unstable state without any phase transition. We expect that such instability may remove under effect of higher order corrections of the entropy discussed in the previous section.\\
Exploiting relations (\ref{ent}) and (\ref{tem}), the higher order corrected entropy (\ref{corrected-entropy}) of Schwarzschild-Beltrami-de Sitter black hole is given by,
\begin{eqnarray}\label{s}
S_i=\pi x_i^2 - {\alpha_1}  \log \frac{l^2-3x_i^2}{\sqrt{\pi}l^2}+2\alpha_1 \log 2 +
 \frac{\alpha_2}{\pi x_i^2},
\end{eqnarray}
where $\alpha_1$ and $\alpha_2$ are dimensionfull parameters.\\
Here, other higher order sub-leading terms are negligibly small, so we have neglected them.
Plugging the value of $x_i$, it is given by
\begin{eqnarray}\label{s1}
S_i= \frac{\pi l^2r_i^2}{l^2+r_i^2-t^2} - {\alpha_1}  \log \frac{l^2-2r_i^2-t^2}{\sqrt{\pi}(l^2+r_i^2-t^2)}+2\alpha_1 \log 2 +
 \frac{\alpha_2(l^2+r_i^2-t^2)}{\pi l^2r_i^2},
\end{eqnarray}
We find that the horizon entropy $S_{h}$ is increasing function of horizon radius in absence of higher order correction. Fig. \ref{fig1} shows that effect of logarithmic entropy with positive $\alpha_{1}$ and negative $\alpha_{2}$ is increasing entropy. Blue dash dot and blue long dash lines of Fig. \ref{fig1} show that negative $\alpha_{1}$ and positive $\alpha_{2}$ violates the second law of thermodynamics and one can obtain negative entropy or decreasing entropy in some regions. Hence similar to the results of the Ref. \cite{sur} we find that $\alpha_{1}$ should be positive and $\alpha_{2}$ should be negative. Solid red line represents the case of only logarithmic correction ($\alpha_{2}=0$), while space dash red line represent only the second order correction ($\alpha_{1}=0$). Effect of both corrections illustrated by dash green line. Later, we will use these result to extract heat capacity and discuss about stability of Schwarzschild-Beltrami-de Sitter black hole.\\

Now, we can study the first law of thermodynamics (\ref{first}) and find the following condition to have validity of this law,
\begin{equation}\label{condition-first law}
\frac{\alpha_{1}}{\alpha_{2}}=\frac{2(l^2+r_i^2-t^2)^{2}}{\pi l^{2} r_{i}^{2}\left((l^2+4r_i^2-t^2)\log{\left(\frac{l^2-2r_i^2-t^2}{2\sqrt{\pi}(l^2+4r_i^2-t^2)}\right)}+\frac{3}{2}r_{i}^{2}\right)}.
\end{equation}
This condition will satisfy with positive $\alpha_{1}$ and negative $\alpha_{2}$ in allowed region of the event horizon radius.\\
Finally, we can study enthalpy which interpreted as the black hole mass \cite{Dolan, JJP}. Hence, using the relation (\ref{mass}) and condition (\ref{condition-first law}) we can write the following expression for the enthalpy,
\begin{eqnarray}\label{enthalpy}
H&=&\frac{(l^{2}-t^{2})r}{2(l^2+r_h^2-t^2)^{\frac{3}{2}}}\nonumber\\
&-& \frac{\alpha_1(l^2-2r_h^2-t^2)}{4\pi r_{h} l(l^2+r_h^2-t^2)^{\frac{3}{2}}}  \left((l^2-2r_h^2-t^2)\log \left( \frac{l^2-2r_h^2-t^2}{4\sqrt{\pi}(l^2+r_h^2-t^2)}\right)-3r_{h}^{2}\right).
\end{eqnarray}
We can see that negative $\alpha_{1}$ yields to negative enthalpy which is corresponding to negative mass which is not physical situation, hence we again see that positive $\alpha_{1}$ is reasonable. At the large $r_{h}$ we can see that corrected and uncorrected results are the same as explained already, and importance of higher order corrections observed for the small black holes. In the next section we have more focus on the free energies.

\section{Gibbs and Helmholtz Free Energies}
\begin{figure}[h!]
 \begin{center}$
 \begin{array}{cccc}
\includegraphics[width=70 mm]{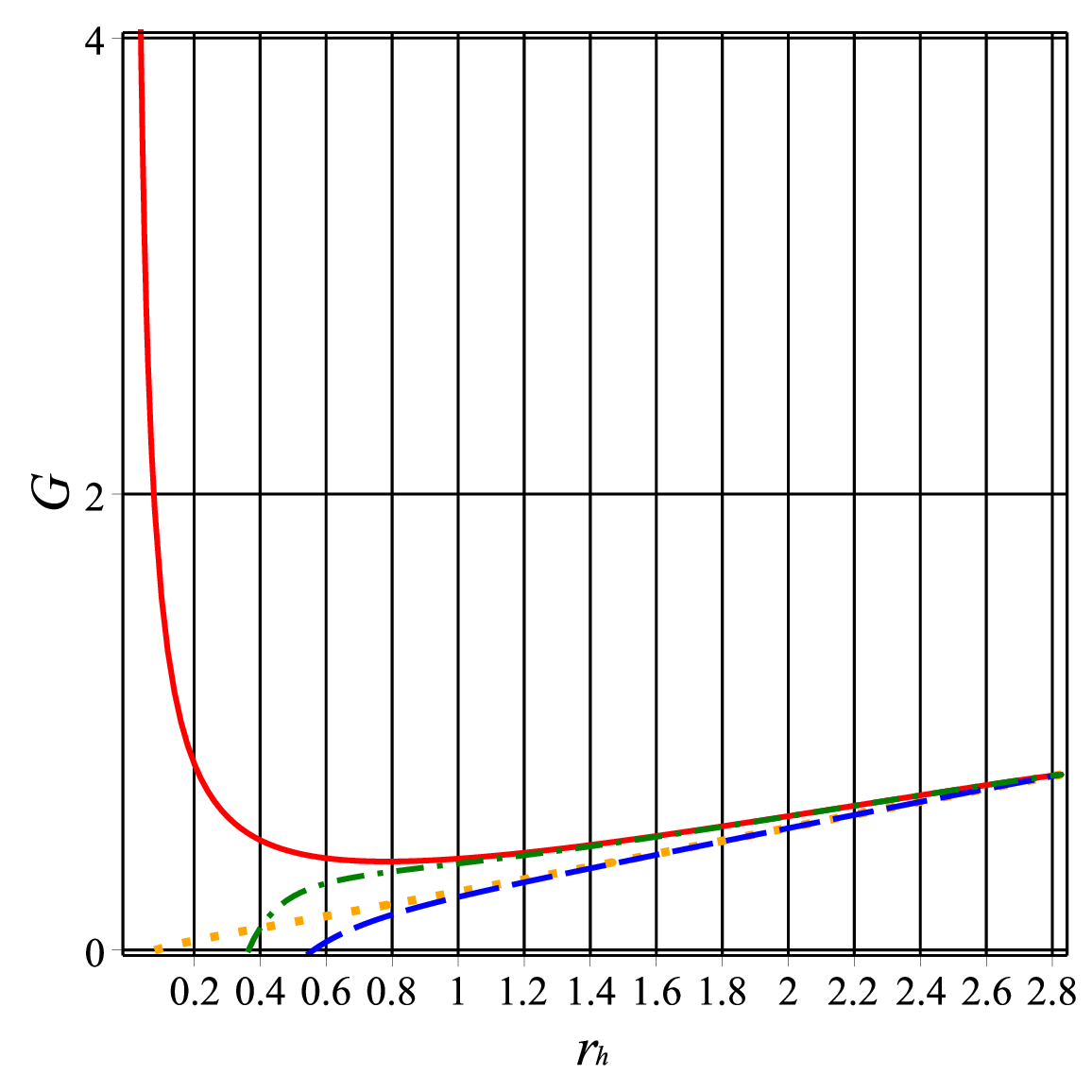}
 \end{array}$
 \end{center}
\caption{Gibbs free energy in terms of $r_{h}$ with $l=4$ and $t=0$. $\alpha_{1}=\alpha_{2}=0$ (dot orange), $\alpha_{1}=1$, $\alpha_{2}=0$ (solid red), $\alpha_{1}=0$, $\alpha_{2}=-1$ (dash blue), $\alpha_{1}=1$, $\alpha_{2}=-1$ (dash dot green).}
 \label{fig2}
\end{figure}

Gibbs free energy is one of the important thermodynamics quantities which can be used to investigate small/large black hole phase transition by computing minimum of $G-T$ diagram at constant pressure.
The Gibbs free energy is defined as \cite{ha}
\begin{eqnarray}
G=M-T_i S_i = T_iS_i-2PV_i,
\end{eqnarray}
Utilizing the expressions of temperature (\ref{tem}), entropy (\ref{s}), volume (\ref{vol}) and pressure (\ref{pr}),
the higher order corrected Gibbs free energy for Schwarzschild-Beltrami-de Sitter black hole is computed by
\begin{eqnarray}
G&=&\frac{x_i(l^2-3x_i^2)}{4l^2} +\frac{x_i^3}{l^2}-\alpha_1\frac{l^2-3x_i^2}{4\pi l^2 x_i}\log \frac{l^2-3x_i^2}{\sqrt{\pi} l^2} +\alpha_1 \left(\frac{\log 2}{2\pi}\right)\frac{l^2-3x_i^2}{l^2x_i}\nonumber\\
&+&\frac{\alpha_2}{4\pi^2}\frac{l^2-3x_i^2}{l^2x_i^3}.
\end{eqnarray}
This is further simplifies to
\begin{eqnarray}\label{Gibbs}
G&=& \frac{r_hl(l^2+2r_h^2-t^2)}{4(l^2+r_h^2-t^2)^{\frac{3}{2}}}-\alpha_1\frac{(l^2-2r_h^2-t^2)}{4\pi lr_h\sqrt{l^2+r_h^2-t^2}}
\log \frac{l^2-2r_h^2-t^2}{\sqrt{\pi}(l^2+r_h^2-t^2)}\nonumber\\
&+& \alpha_1\frac{\log 2}{2\pi}\frac{(l^2-2r_h^2-t^2)}{lr_h\sqrt{l^2+r_h^2-t^2}}+
\frac{\alpha_2}{4\pi^2}\frac{(l^2-2r_h^2-t^2)(l^2+r_h^2-t^2)^\frac{1}{2}}{r_h^3l^3}.
\end{eqnarray}
In the Fig. \ref{fig2} we can see the effect of correction terms on the Gibbs free energy by variation of horizon radius. It is clear that logarithmic correction enhanced Gibbs free energy (see solid red line of Fig. \ref{fig2}), while second order correction decreases its value (see dash blue line of Fig. \ref{fig2}). When we consider both corrections (dash dot green line of Fig. \ref{fig2}), there is a critical radius $r_{c}$ then value of Gibbs free energy enhanced for $r_{h}>r_{c}$ while decreased for $r_{h}<r_{c}$. For selected values $t=0$ and $l=4$ we found $r_{c}\approx0.4$. It means that the Schwarzschild-Beltrami-de Sitter black hole with the second order correction ($\alpha_{1}=0$ and $\alpha_{2}\neq0$) is more stable because have smallest Gibbs free energy. We can also remove $\alpha_{2}$ by using the condition (\ref{condition-first law}) and find similar behavior.\\

\begin{figure}[h!]
 \begin{center}$
 \begin{array}{cccc}
\includegraphics[width=70 mm]{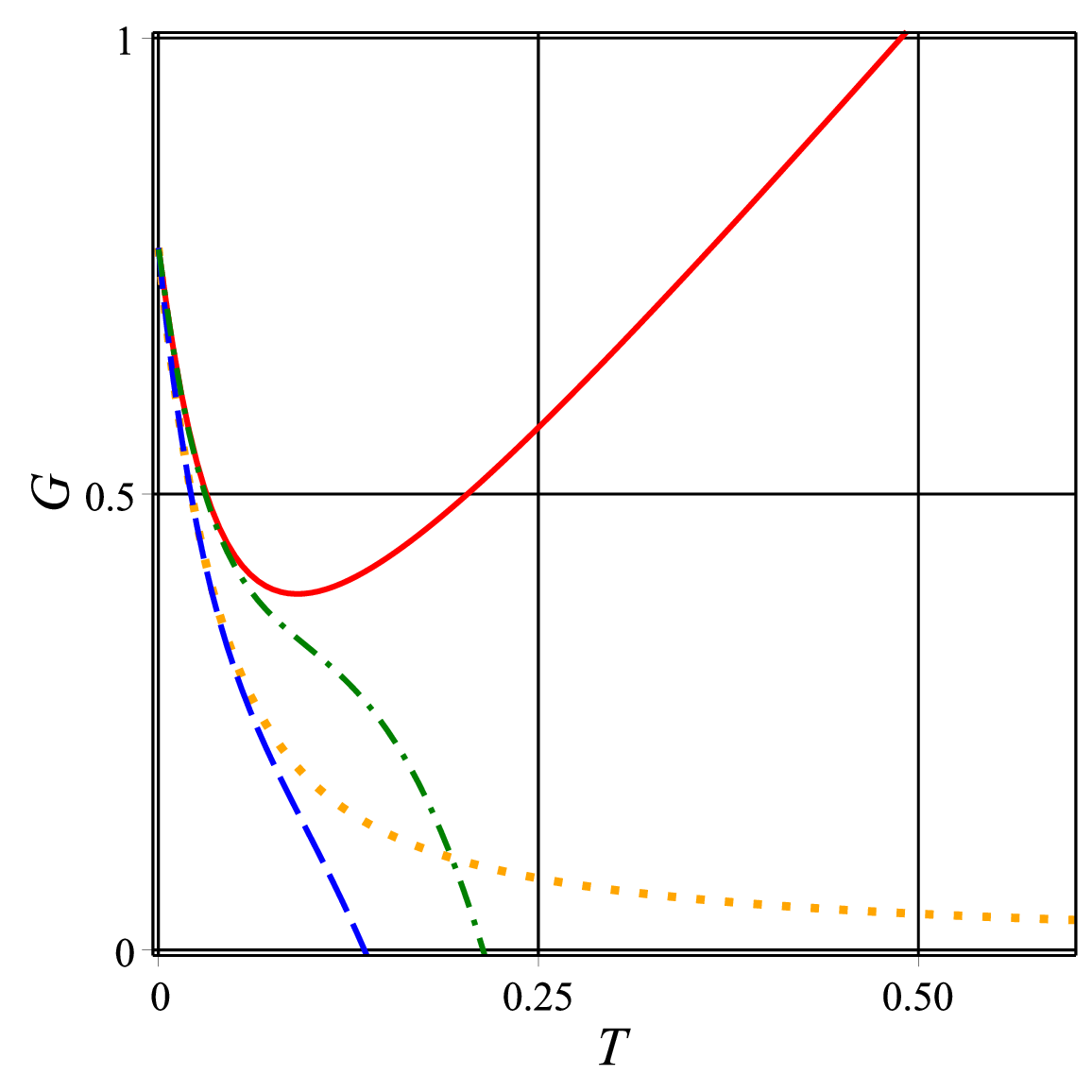}
 \end{array}$
 \end{center}
\caption{Gibbs free energy in terms of $T$ with $l=4$ and $t=0$. $\alpha_{1}=\alpha_{2}=0$ (dot orange), $\alpha_{1}=1$, $\alpha_{2}=0$ (solid red), $\alpha_{1}=0$, $\alpha_{2}=-1$ (dash blue), $\alpha_{1}=1$, $\alpha_{2}=-1$ (dash dot green).}
 \label{fig3}
\end{figure}

By using the relation (\ref{tem}) one can obtain horizon radius versus temperature, hence we can rewrite Gibbs free energy in terms of temperature (result illustrated by the Fig. \ref{fig3}). As discussed by the Ref. \cite{ha} there is no critical point and phase transition in absence of corrections. But from solid red line of the Fig. \ref{fig3} we can see a minimum of Gibbs free energy which shows an equilibrium state at constant pressure. It is clear that the corrected Gibbs energy has different behavior with the uncorrected one.\\

For, Schwarzschild-de Sitter black hole, the higher order corrected Gibbs free energy
has following form,
\begin{eqnarray}\label{Gibbs-sd}
G_{sd}&=& \frac{r_h(l^2-r_h^2)}{4l^2}+\frac{r_h^3}{ l^2} -\alpha_1\frac{(l^2-r_h^2)}{4\pi l^2r_h}
\log \frac{l^2-r_h^2}{\sqrt{\pi}l^2}\nonumber\\
&+& \alpha_1\frac{\log 2}{2\pi}\frac{(l^2-r_h^2)}{l^2r_h}+
\frac{\alpha_2}{4\pi^2}\frac{(l^2-r_h^2)}{r_h^3l^2}.
\end{eqnarray}

\begin{figure}[h!]
 \begin{center}$
 \begin{array}{cccc}
\includegraphics[width=70 mm]{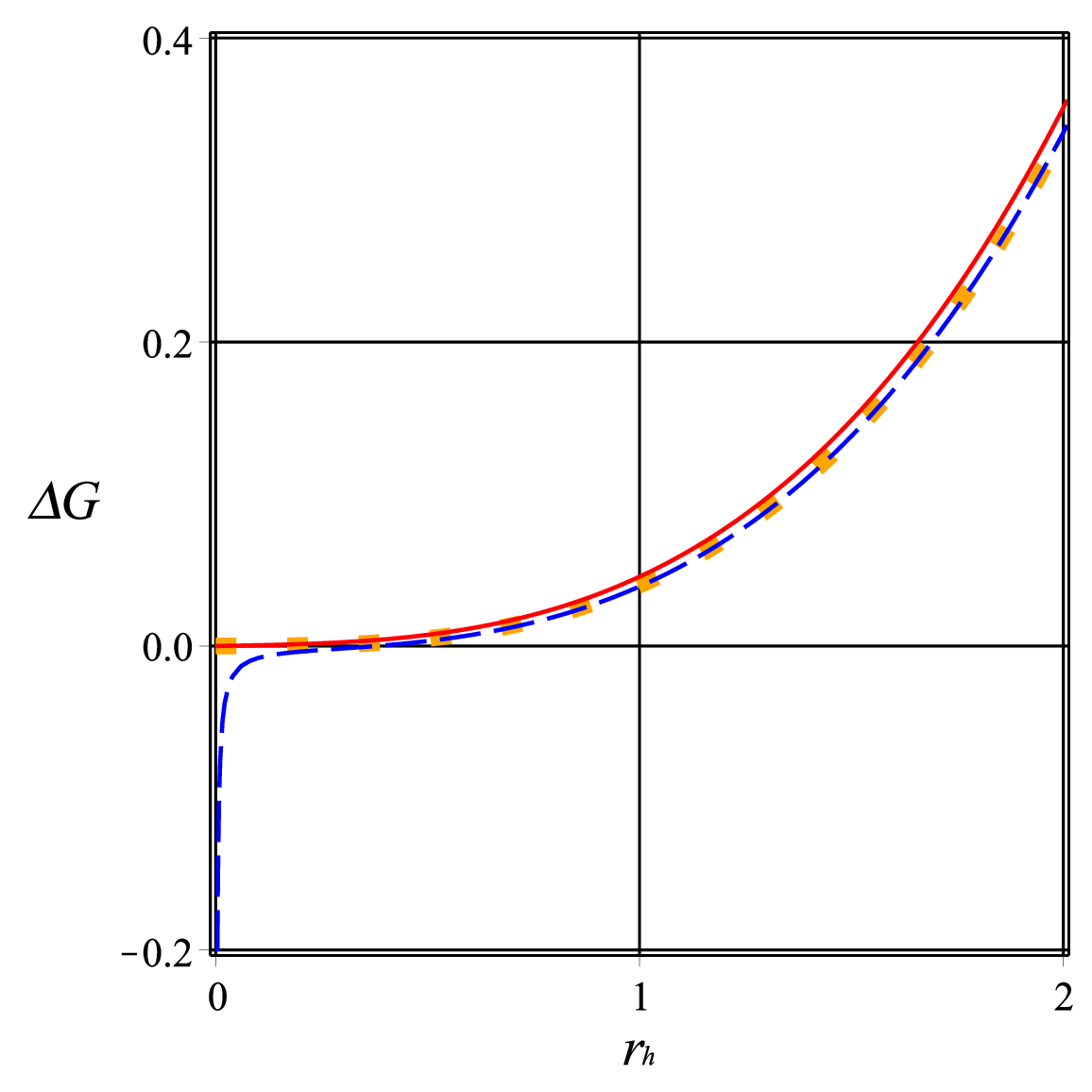}
 \end{array}$
 \end{center}
\caption{$\Delta G$ in terms of $r_{h}$ with $l=4$ and $t=0$. $\alpha_{1}=\alpha_{2}=0$ (dot orange), $\alpha_{1}=1$, $\alpha_{2}=0$ (solid red), $\alpha_{1}=0$, $\alpha_{2}=-1$ (dash blue).}
 \label{fig4}
\end{figure}

Hence, we can study difference between Gibbs free energy of the Schwarzschild-Beltrami-de Sitter black hole given by (\ref{Gibbs}) and the Schwarzschild-de Sitter black hole given by (\ref{Gibbs-sd}),
\begin{equation}\label{Gibbs-delta}
\Delta G=G_{sd}-G.
\end{equation}
As discussed in Ref. \cite{ha} it is positive for $\alpha_{1}=\alpha_{2}=0$ which means the Schwarzschild-Beltrami-de Sitter black hole is more stable than the Schwarzschild-de Sitter black hole. In presence of the logarithmic correction we find similar result (see soled red line of the Fig. \ref{fig4}). However, in presence of the second order correction (blue dashed line of the Fig. \ref{fig4}) we can see that $\Delta G$ is negative for $r_{h}<r_{c}$ which means that the second order corrected of Schwarzschild-de Sitter black hole may be more stable than the Schwarzschild-Beltrami-de Sitter black hole with the second order correction of the entropy.

Gibbs free energy can be used to obtain Helmholtz free energy $F$ via the following relation,
\begin{equation}\label{Helmholtz}
F=G-PV.
\end{equation}
By using the condition (\ref{condition-first law}), $t=0$, and $l=4$ we can obtain the following Helmholtz free energy,
\begin{eqnarray}\label{Helmholtz2}
F&=&\frac{(512+64r_{h}^{2}+r_{h}(r_{h}^{2}+16)^{\frac{3}{2}})r_{h}}{32(r_{h}^{2}+16)^{\frac{3}{2}}}\nonumber\\
&+& \alpha_1\frac{48-6r_{h}^{2}-(16-2r_{h}^{2})^{2}\log{\left(\frac{16-2r_{h}^{2}}{4\sqrt{\pi}(16+r_{h}^{2})}\right)}}{32\pi r_{h}(r_{h}^{2}+16)^{\frac{3}{2}}}.
\end{eqnarray}
In presence of the logarithmic correction there is a minimum for the Helmholtz free energy. Vanishing $\alpha_{1}$ gives the Helmholtz free energy of the Schwarzschild-Beltrami-de Sitter black hole as increasing function of horizon radius without any extremum.\\
In the next section we extract heat capacity and discuss about stability of the system under effect of higher order corrections of the entropy.

\section{Heat Capacity and Stability}
\begin{figure}[h!]
 \begin{center}$
 \begin{array}{cccc}
\includegraphics[width=70 mm]{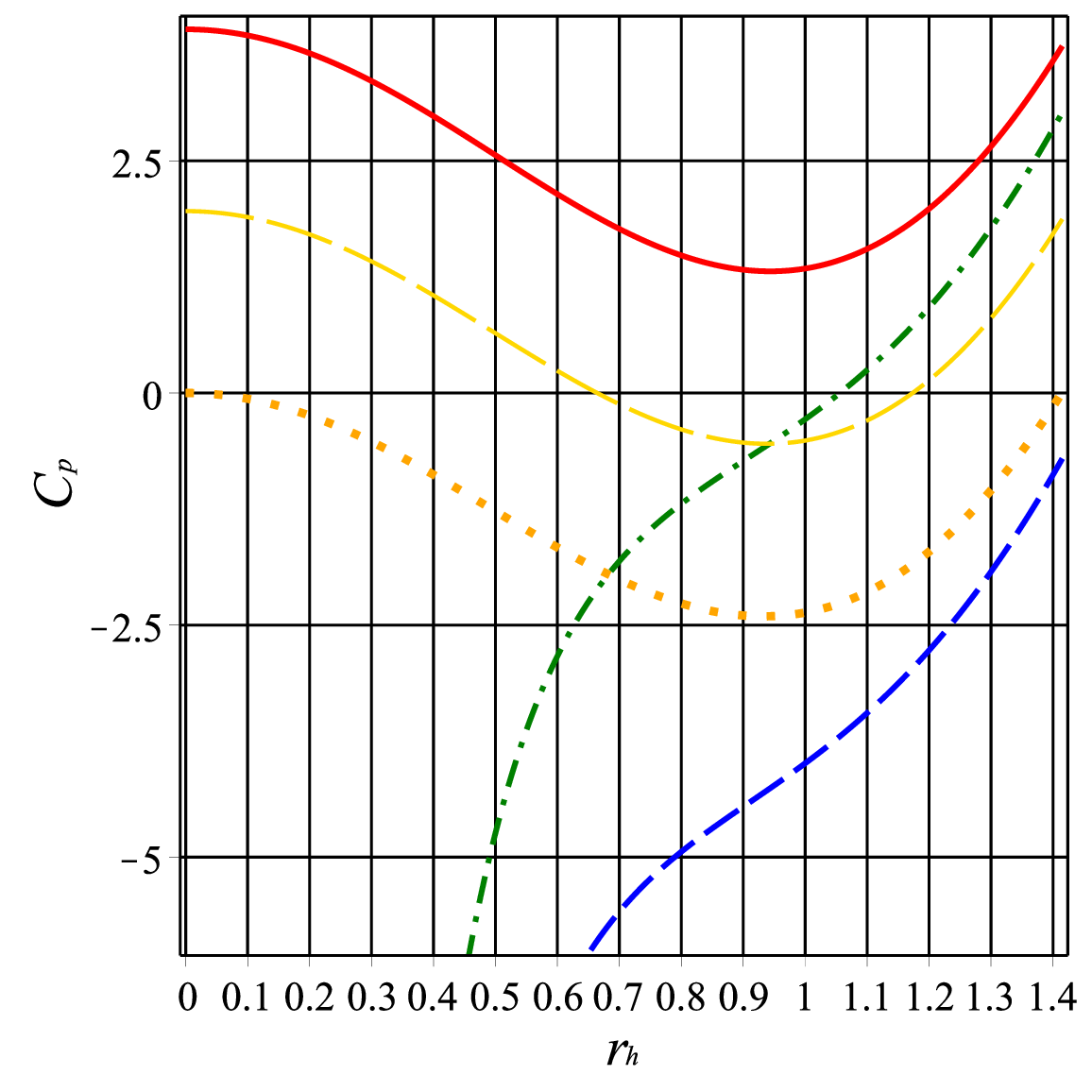}
 \end{array}$
 \end{center}
\caption{Heat capacity in terms of $r_{h}$ with $l=4$ and $t=0$. $\alpha_{1}=\alpha_{2}=0$ (dot orange), $\alpha_{1}=1$, $\alpha_{2}=0$ (solid red), $\alpha_{1}=0.5$, $\alpha_{2}=0$ (long dash gold), $\alpha_{1}=0$, $\alpha_{2}=-1$ (dash blue),  $\alpha_{1}=1$, $\alpha_{2}=-1$ (dash dot green).}
 \label{fig5}
\end{figure}

Heat capacity or specific heat is one of the important thermodynamics quantity which will be used to find stability state of the black hole. If heat capacity be positive, then the black hole is in stable phase, and divergence of heat capacity shows phase transition.
The heat capacity in constant pressure is defined by,
\begin{eqnarray}
C_P=\frac{\partial M}{\partial r_h}\left(\frac{\partial T_h}{\partial r_h}\right)^{-1}.
\end{eqnarray}
For the Schwarzschild-Beltrami-de Sitter black hole with the corrected entropy at constant pressure, it is calculated by
\begin{eqnarray}
C_P =
 && \frac{2}{  \pi   l^2 r_h^2 \left(l^2+r_h^2-t^2\right)(l^2+4r_h^2-t^2)}  \left[\pi^2l^4r_h^4(8r_h^2 +t^2-l^2)  -4\pi
  \alpha_1  l^4 r_h^2 t^2 \log 2\right. \nonumber\\
 &&-\left. \pi  \alpha_1  l^4 r_h^2 t^2 \log  \pi  -6 \pi  \alpha_1  l^2 r_h^6+8 \pi  \alpha _1 l^2 r_h^6 \log  2 +6 \pi  \alpha_1  l^2 r_h^4 t^2-10 \pi  \alpha_1  l^2 r_h^4 t^2 \log  2
 \right. \nonumber\\
 &&\left. +2\pi  \alpha_1  l^2 r_h^2 t^4 \log 2
 +2 \pi  \alpha_1  l^4 r_h^2 t^2 \log \left(\frac{l^2-2 r_h^2-t^2}{l^2+r_h^2-t^2}\right)+2\pi  \alpha_1  l^6 r_h^2 \log  2 -6 \pi  \alpha_1  l^4 r_h^4 \right. \nonumber\\
 &&\left. -\pi  \alpha_1  l^2 r_h^2 \left(l^4+5 l^2 r_h^2+4 r^4-5 r_h^2 t^2+t^4\right) \log \left(\frac{l^2-2 r_h^2-t^2}{\sqrt{\pi } \left(l^2+r_h^2-t^2\right)}\right) +10 \pi  \alpha_1  l^4 r_h^4 \log  2  \right. \nonumber\\
 &&\left.       +3 \alpha_2  l^6+6 \alpha_2  l^4 r_h^2 -9 \alpha_2  l^4 t^2+3 \alpha_2  l^2 r_h^4  -3 \alpha_2   t^6 \right. \nonumber\\
 &&\left. -12 \alpha_2   l^2 r_h^2 t^2  +9 \alpha_2   l^2 t^4-3 \alpha_2   r_h^4 t^2+6 \alpha_2   r_h^2 t^4  \right]
\end{eqnarray}
In the Fig. \ref{fig5} we can see behavior of heat capacity at constant pressure of the Schwarzschild-Beltrami-de Sitter black hole under effect of thermal fluctuation until the second order. As discussed in Ref. \cite{ha}, vanishing thermal fluctuations yields to completely negative heat capacity in allowed regions (see orange dotted line of the Fig. \ref{fig5}) and the Schwarzschild-Beltrami-de Sitter black hole is unstable. Significant result is that if we switch on logarithmic correction with appropriate value (like $\alpha_{1}=1$), then heat capacity is completely positive and the Schwarzschild-Beltrami-de Sitter black hole is stable (see red solid line of the Fig. \ref{fig5}). However, with other values of $\alpha_{1}$ we can have stable to unstable, and also unstable to stable phase transition (see golden long dashed line of the Fig. \ref{fig5}). On the other hand if only switch on the second order correction we have still unstable Schwarzschild-Beltrami-de Sitter black hole (see blue dashed line of the Fig. \ref{fig5}) which means that the last term of the equation (\ref{corrected-entropy}) has no any effect in stability of black hole. Finally, in presence of both corrections we can see unstable to stable phase transition, for example at $r_{h}\approx 1.05$ for $l=4$, $t=0$, $\alpha_{1}=1$ and $\alpha_{2}=-1$ (see green dash dotted line of the Fig. \ref{fig5}). The case of negative $\alpha_{2}$ is completely agree with arguments of the Ref. \cite{sur}.\\

An important consequence is that change in coordinates from inertial to non-inertial reference drastically changes the well-understood thermodynamic
properties of the metric. However, we can use higher order correction of the entropy to reproduce well-known thermodynamics such as thermodynamics stability.

\section{Conclusions}
In this paper, we considered Schwarzschild-Beltrami-de Sitter black hole and studied thermodynamics in presence of higher order corrections of the entropy, specially the first (logarithmic) and the second orders. Approximately we assume that Schwarzschild-Beltrami-de Sitter black hole is static, hence we are able to consider usual thermodynamics relations. We used results of the Ref. \cite{sur} for the corrected entropy where the first order correction is logarithmic while the second order correction is proportional to inverse of original entropy with negative coefficient. However we considered general coefficient for the first and second orders and confirmed  results of the Ref. \cite{sur}. By using thermodynamical analysis, we found appropriate condition to satisfy the first law of thermodynamics. Also we found that the corrected entropy as well as original entropy is increasing function and hence satisfy the second law of thermodynamics. We obtained Gibbs and Helmholtz free energies and found effects of thermal fluctuations of the second order. We found that logarithmic correction enhanced Gibbs free energy (see solid red line of Fig. \ref{fig2}), while second order correction decreases its value. We have shown that the effect of logarithmic correction on the Gibbs free energy is arising a minimum, we also found that the second order corrected of Schwarzschild-de Sitter black hole may be more stable than the Schwarzschild-Beltrami-de Sitter black hole with the second order correction of the entropy.\\
It has been argued that Schwarzschild-Beltrami-de Sitter black hole is unstable without any phase transition and critical points \cite{ha}. By using behavior of the heat capacity we studied the effect of the logarithmic correction in stability of Schwarzschild-Beltrami-de Sitter black hole. We have shown that presence of the first- and second-order corrections of the entropy yield to unstable/stable phase transition, and led to stability of black hole at the allowed region of the event horizon. Finally it would be interesting to consider the possibility that similar results work also in most general theories. For example it is interesting to discuss anti evaporation Schwarzschild-Beltrami-de Sitter black hole and study such as \cite{cop1} in the context of extended theories of  gravity \cite{cop2}.

\end{document}